\documentclass[%
reprint,
superscriptaddress,
pra,
]{revtex4-2}

\usepackage{graphicx} % Required for inserting images
\usepackage{amsmath} 
\usepackage{xcolor}
\usepackage{braket}
\usepackage[colorlinks=true, linkcolor=red, anchorcolor=black, citecolor=blue,
urlcolor = blue]{hyperref}
\usepackage[autostyle]{csquotes}
\newcommand{\mycomment}[1]{}
\usepackage{soul}

\begin{document}

\title{Observation of Asymmetric Sideband Generation in Strongly-driven Rydberg Atoms}

\author{Dangka Shylla}
 \affiliation{Department of Physics, University of Colorado, Boulder, Colorado 80302, USA}
\affiliation{Associate of the National Institute of Standards and Technology, Boulder, Colorado 80305, USA}
\author{Nikunjkumar Prajapati}
\affiliation{ National Institute of Standards and Technology, Boulder, Colorado 80305, USA}
\author{Andrew P. Rotunno}
\affiliation{ National Institute of Standards and Technology, Boulder, Colorado 80305, USA}
\author{Noah Schlossberger}
\affiliation{ National Institute of Standards and Technology, Boulder, Colorado 80305, USA}
%\affiliation{Department of Physics, University of Colorado, Boulder, Colorado 80302, USA}
%\affiliation{Associate of the National Institute of Standards and Technology, Boulder, Colorado 80305, USA}
\author{Dixith Manchaiah}
\affiliation{Department of Physics, University of Colorado, Boulder, Colorado 80302, USA}
\affiliation{Associate of the National Institute of Standards and Technology, Boulder, Colorado 80305, USA}
\author{William J. Watterson}
\affiliation{Department of Physics, University of Colorado, Boulder, Colorado 80302, USA}
%\affiliation{ National Institute of Standards and Technology, Boulder, Colorado 80305, USA}
\affiliation{Associate of the National Institute of Standards and Technology, Boulder, Colorado 80305, USA}
\author{Alexandra Artusio-Glimpse}
\affiliation{ National Institute of Standards and Technology, Boulder, Colorado 80305, USA}
\author{Samuel Berweger}
\affiliation{ National Institute of Standards and Technology, Boulder, Colorado 80305, USA}
\author{Matthew T. Simons}
\affiliation{ National Institute of Standards and Technology, Boulder, Colorado 80305, USA}
\author{Christopher L. Holloway}
\affiliation{ National Institute of Standards and Technology, Boulder, Colorado 80305, USA}
 
\date{\today}

\begin{abstract}
Improving the bandwidth of Rydberg atom-based receivers is an ongoing challenge owing to the long-lived Rydberg state lifetimes that limit the refresh rate of ground state atoms.
In particular, the LO-based Rydberg mixer approach allows for bandwidths into the few-MHz range. 
Here, we use heterodyne detection of the Rydberg atom receiver probe laser to separate the negative and positive sidebands that originate from distinct six wave mixing processes, in order to investigate their individual bandwidths.
We experimentally confirm the prediction that the negative sideband exhibits a higher bandwidth than the positive sideband.
We further explore the effect of coupling and probe laser Rabi frequency on the bandwidth, which we find to be in good agreement with our model. We achieved a maximum experimental (and theoretical) bandwidth of about 11 (11)~MHz and 3.5 (5)~MHz for the negative and positive sidebands, respectively, from the -3dB roll-off point for optimized field parameters.
This work provides insight into the bandwidth-limiting features of Rydberg atom receivers and points the way towards further optimization of their response.
\end{abstract}

\maketitle

\section{Introduction}\label{sec:intro}
Excited Rydberg atomic states are sensitive to microwave-band fields due to strong electric dipole moments between nearby bound atomic states \cite{sedlacek2012microwave,6910267,9748947}.
These Rydberg-atom sensors can receive time-varying signals while performing demodulation and down-conversion automatically.
As such, Rydberg sensors have been used in applications ranging from radio frequency (RF) to terahertz regime, %\textcolor{red}{XXX to XXX},
% many have realized applications of broadband reception of automatically down-converted \cite{simons2019rydberg} electric field signals \cite{meyer2020assessment},
%\cite{song1,meyer1,cox1,holl4,anderson2,10.1063/5.0048415,10268327,
%doi:10.1063/1.5099036,8778739,meyer2020assessment,doi:10.1116/5.0098057}, 
notably breaking the Chu limit for small receivers \cite{cox2018quantum}. 
There has been notable interest in the development of Rydberg sensors and current thrusts are geared towards improving the sensitivity and bandwidth.
In this work we focus on the bandwidth, which is in general limited by the kHz decay rates of the Rydberg states that limit the refresh rate of ground state atoms available for measurement. 

In order to provide the phase sensitive detection needed to demodulate digital communication signals \cite{song1, meyer1, cox1, holl4, anderson2, Otto21}, Rydberg sensors are often used in a \enquote{Rydberg mixer} configuration where a local oscillator (LO) is broadcast onto the atoms at a frequency near the signal of interest in order to generate a down-converted beat note signal that carries the phase and amplitude of the signal detected via Autler-Townes (AT) splitting induced by the LO in the electromagnetically induced transparency (EIT) spectrum \cite{simons2019rydberg,jing2020atomic}.
It was recently proposed that the measured mixer signal is the result of two separate sidebands shifted to either side of the laser frequency due to two distinct six-wave mixing processes \cite{firdoshi2022six,yang2023high}.
It was further suggested that the two sidebands exhibit distinct instantaneous bandwidths, though this is difficult to verify experimentally since the standard self-homodyne probe laser Rydberg mixer detection scheme does not allow experimental access to the individual sidebands.

In this work, we use a heterodyne detection scheme in order to separate the two sidebands in the frequency domain and study their individual bandwidth characteristics. 
Our measurements confirm that the negative sideband exhibits an overall higher instantaneous bandwidth than the positive one.
We further investigate the effect of the Rabi frequency of the probe and coupling laser fields on bandwidth, where we find that higher coupling Rabi frequency increases the bandwidth whereas higher probe Rabi frequency decreases the bandwidth.  %\textcolor{red}{XXX}.
We support our results with modeling, where we find good quantitative agreement with the experimentally measured bandwidths for lower values of the probe Rabi frequency.
These results provide a clear path for improving the achievable Rydberg sensor bandwidth both in terms of the underlying wave mixing mechanisms as well as Rabi frequencies of the optical fields.

\section{Energy Level Scheme and Experimental setup}\label{sec:exp}
%%%%%%%%%%%%%%%%%%%%%%%%%%%Figure1%%%%%%%%%%%%%%%%%%%%%%%%%
\begin{figure}
    \centering
\includegraphics[width=1.02\columnwidth]{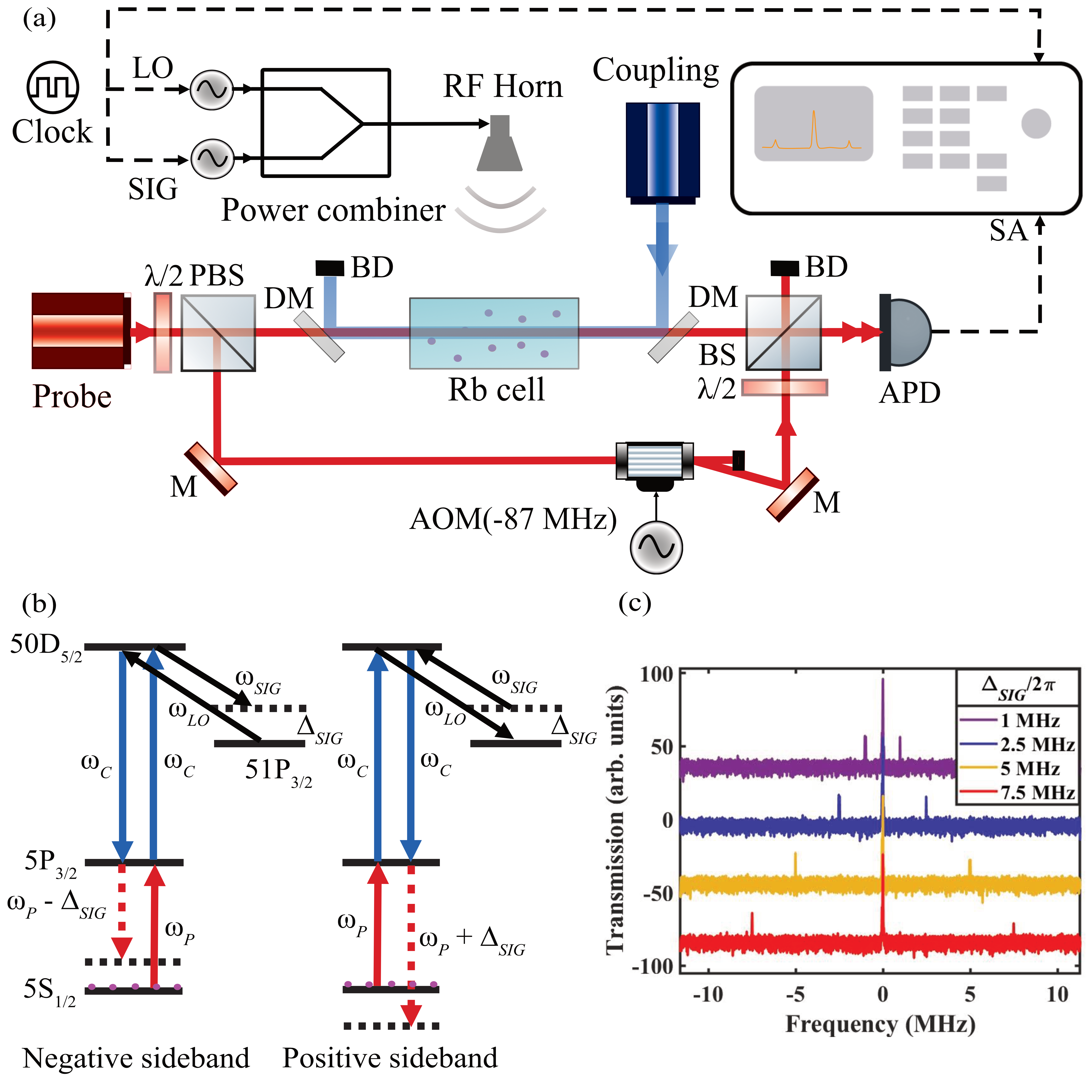}
    \caption{  (a) Experimental setup using a heterodyne reference beam frequency shifted using an AOM operating at $\omega_{AOM}=$ 87~MHz to differentiate between the negative and positive amplitude modulation sidebands.  
    (b) Energy level diagrams of the four-level Rydberg EIT-AT scheme in $^{85}$Rb employing the LO-based Rydberg mixer configuration to generate the negative ($\omega_{P}-\Delta_{SIG}$) and positive ($\omega_{P}+\Delta_{SIG}$) sidebands on the probe transmission, respectively.
    %\textcolor{red}{Need clear labels on diagrams to indicate which one corresponds to + and - sidebands; text in d is too small to read well}
    (c) Spectrum analyzer data of the probe transmission for various RF signal field detuning, $\Delta_{SIG}/(2\pi)$= 1, 2.5, 5 and 7.5~MHz (from top to bottom), keeping the frequency of LO fixed. Here, the central frequency corresponds to $\omega_{AOM}$ that has been rescaled for simplicity. $\lambda/2$: Half waveplate; PBS: Polarising beam splitter; BS: Beam splitter;  DM: Dichoric mirror; APD: Avalanche photodetector; BD: Beam dump;  SA: Spectrum analyser.}
    \label{fig:scheme}
\end{figure}
%%%%%%%%%%%%%%%%%%%%%%%%%%%%%%%%%%%%%%%%%%%%%%%%%%%%%%%%%%%%

The experimental setup corresponding to our scheme is shown in Figure \ref{fig:scheme}(a), where a 780~nm probe laser and a 480~nm coupling laser are counter-propagating within a Rb vapor cell at room temperature.  
The $1/e^2$ radii of the probe and coupling beams are kept fixed at 148~$\mu$m and 180~$\mu$m, respectively.
A key feature of our setup is the use of a 780~nm reference beam that is frequency shifted using an acousto-optic modulator (AOM) by $\omega_{AOM}=$ 87~MHz and is made to interfere with the transmitted probe beam. 
This method of optical heterodyning of the probe beam shifts the beat note signal by $\omega_{AOM}$ to separate the negative and the positive amplitude modulation sidebands in frequency such that when we apply the RF signal field detuning, $\Delta_{SIG}$ we see sidebands generated on either side of $\omega_{AOM}$.

The energy level diagram of the four-level Rydberg EIT-AT scheme in $^{85}$Rb is shown in Figure \ref{fig:scheme}(b). 
The probe laser ($\omega_P$) at 780~nm and the coupling laser ($\omega_C$) at 480~nm drive the $5S_{1/2}$ to $5P_{3/2}$ and  $5P_{3/2}$ to $50D_{5/2}$ transitions, respectively. 
The probe transmission is detected using an avalanche photodetector (APD) with a bandwidth of 400~MHz and recorded using a spectrum analyzer (SA).
Alongside these lasers, there are two RF fields applied to generate a Rydberg mixer: the LO field ($\omega_{LO}$) is resonant on the  $50D_{5/2}$ to $51P_{3/2}$ Rydberg transition at 17.041~GHz and the weak signal field ($\omega_{SIG}$) is at a frequency difference of $\Delta_{SIG}$ relative to the LO field.
The two RF fields are sourced from signal generators and are combined using a power combiner before being directed to an RF horn. 
Both RF fields are incident perpendicular to the probe and coupling lasers and are co-polarized with these. 
The presence of the LO field is typically used to generate a beat note signal at the frequency difference between these two fields that can be read out at frequency $\Delta_{SIG}$ from the oscillating probe transmission.

% This beat signal represents the sum of the positive and negative amplitude modulation sidebands of the oscillating probe transmission, which would appear at the same frequency in a spectrum analyzer trace. This phenomenon has been attributed to a six-wave mixing process \cite{yang2023high}. 
 %The lifetime of the first excited state $5P_{3/2}$ is 26.25 ns \cite{SAF11}  and the effective  lifetime of the Rydberg states $50D_{5/2}$ and $51P_{3/2}$ are 65.352 $\mu$s and 86.547 $\mu$s respectively \cite{BET09}. The dipole moment matrix element for  $5S_{1/2}$ to $5P_{3/2}$ and  $5P_{3/2}$ to $50D_{5/2}$ transitions are 1.95696, 0.01233968, 1574.69135769 and 3214.32527392(ea0) respectively
 
Figure \ref{fig:scheme}(c) shows the negative and positive sideband spectra at indicated values of the RF signal field detuning, $\Delta_{SIG}/(2\pi)$ in MHz.
The central peak is centered at $\omega_{AOM}=$ 87~MHz due to the heterodyne detection, and two sideband peaks are clearly seen.
In the case of the standard self-homodyne measurement typically performed in Rydberg sensing, these two sidebands are overlapped and are thus indistinguishable. 
These sidebands have previously been attributed to a six-wave mixing process \cite{yang2023high} and their origin is schematically shown in \ref{fig:scheme}(b).

% The 780~nm probe laser is generated from an external cavity diode laser (ECDL) and is locked using the Saturated Absorption Spectroscopy (SAS) technique, whereas the 480~nm laser generated from a frequency-doubled diode laser is locked to an EIT signal in a different Rb vapor cell.  %Also in our experiment, we choose the probe laser power of 0.2, 1.2, 4.7, 10.7, 18.9 and 57.8 $\mu$W corresponding to peak Rabi frequencies of $\Omega_P/(2\pi)=1.66, 4.06, 8.03, 16.10 $ and $ 28.17$ MHz respectively, and coupling laser power of 285, and 1200 mW corresponding to peak Rabi frequency of $\Omega_C/(2\pi)=9.87$ and $21.08$ MHz. 
% The two RF fields generated from two RF signal generators with control over frequency and amplitude are combined using a power combiner before being directed to an RF horn. Both RF fields are incident perpendicular to the probe and coupling lasers and are co-polarized with these. 
% The two RF sources and the spectrum analyzer are synchronized using a 10~MHz reference from a Rb clock.  

Our focus here is on the instantaneous bandwidth available for RF signal detection, which underpins achievable data rates.
We measure the signal amplitudes of both sidebands as a function of $\Delta_{SIG}/(2\pi)$ and define the bandwidth as the -3~dB roll-off point.

% Bandwidth can refer to the range of frequencies that can be received by Rydberg states, or the effective rate of change of the radio frequency (RF) field that can be measured. The latter, termed ``instantenous bandwidth,'' will be discussed here. We consider the resonant local oscillator (LO) at optimized power with a near-resonant signal field. 
% We specifically measure the sensor response in terms of the frequency difference between these two signals as the LO is held on resonance and signal is swept away from resonance. % at the same power.  
% The particular fall-off of the bandwidth is worthy of more study, and %\textcolor{red}{
% here we use the -3~dB point %, the -10~dB point, and RLC time constants}
%  to quantify the observed roll-offs.  
% To measure application spaces, parameters such as signal-to-noise (SNR) ratio across modulation rates are used to measure the bandwidth. 
 
\section{Theoretical Model}\label{sec:the}
 
The four-level Rydberg EIT-AT system is theoretically modeled by considering the Hamiltonian H, under the electric-dipole and rotating-wave approximation and in the rotating frame. It is expressed as
\begin{align}
H &= \frac{\hbar}{2}\big[\Omega_{P}\ket{1}\bra{2}+\Omega_{C}\ket{2}\bra{3}+(\Omega_{LO}+\Omega_{SIG}e^{-i\Delta t})\ket{3}\bra{4}\nonumber \\
&\quad -2\Delta_{P}\ket{2}\bra{2}-2(\Delta_{P}+\Delta_{C})\ket{3}\bra{3}\nonumber \\
&\quad-2(\Delta_{P}+\Delta_{C}-\Delta_{LO})\ket{4}\bra{4}+h.c.\big], \nonumber \\
\label{Hami}
\end{align}
where $\ket{1}=5\textrm{S}_{1/2}(\textrm{F}=3)$, $\ket{2}=5\textrm{P}_{3/2}(\textrm{F}=4)$, $\ket{3}=50\textrm{D}_{5/2}$ and $\ket{4}=51\textrm{P}_{3/2}$. $\Omega_P$, $\Omega_C$, $\Omega_{LO}$ and $\Omega_{SIG}$ are the Rabi frequency of the probe, coupling, LO and signal fields with corresponding detunings $ \Delta_{P} $, $ \Delta_{C} $, $ \Delta_{LO} $ and $ \Delta_{SIG} $, respectively. $\Delta=\Delta_{LO}-\Delta_{SIG}$ is the frequency difference between the two RF fields. The generalized Rabi frequency $\Omega_{ij}$ between the states $|i\rangle$ and $|j\rangle$ is defined as $\Omega_{ij} = \frac{E_{ij} \langle i|r|j\rangle}{\hbar}$, where $E_{ij} = \sqrt{\frac{2I}{c \epsilon_0}}$ is the amplitude of the electromagnetic field and $\langle i|r|j\rangle$ is the dipole moment matrix element between these states. The intensity $I$ of the electromagnetic field follows a Gaussian distribution with radial distance $r$ from the beam axis represented by $I = I_0 \exp\left(-\frac{2r^2}{\omega_0^2}\right)$, where $I_0 = \frac{2P}{\pi \omega_0^2}$ is the peak intensity determined by the total power $P$ of the electromagnetic field and the beam waist $\omega_0$. The constants c and $\epsilon_0$  are the speed of light in vacuum and the permittivity of free space, respectively. %The generalised Rabi frequency for the fields is defined as $ \Omega_{ij}=E_{ij}\bra{i}r\ket{j}/\hbar $, where $ E_{ij} $ is the amplitude of the electromagnetic field and $\bra{i}r\ket{j}$ is the dipole moment matrix element  for the $ \ket{i}\rightarrow \ket{j} $ transition.
The dipole moment matrix element for the $ \ket{1}\rightarrow \ket{2} $, $ \ket{2}\rightarrow \ket{3} $ and $ \ket{3}\rightarrow \ket{4} $ transitions are $1.957(ea_0)$, $0.0123(ea_0)$ and $1574.691(ea_0)$, respectively, where $e$ is the charge of the electron and $a_0$ is the Bohr radius. In our experiment, we took measurements with the following probe laser powers: 0.2~$\mu$W, 1.2~$\mu$W, 4.7~$\mu$W, 18.9~$\mu$W, 42.5~$\mu$W and 57.8~$\mu$W which corresponds to peak Rabi frequencies of $\Omega_P/(2\pi)$ = 1.66~MHz, 4.06~MHz, 8.03~MHz, 16.11~MHz, 24.15~MHz and 28.17~MHz, and coupling laser powers: 72~mW, 285~mW and 950~mW which corresponds to peak Rabi frequencies of $\Omega_C/(2\pi)$ = 4.96~MHz, 9.87~MHz and 18.02~MHz, respectively. %$\Omega_P/(2\pi)$=1.17\textcolor{blue}{~MHz}, 2.87\textcolor{blue}{~MHz}, 5.68\textcolor{blue}{~MHz}, 11.39\textcolor{blue}{~MHz}, 17.08\textcolor{blue}{~MHz} and $19.92$~MHz, and coupling laser power of \textcolor{blue}{72~mW}, 285\textcolor{blue}{~mW} and 950~mW which corresponds to \textcolor{blue}{root-mean-square (rms)} Rabi frequencies of $\Omega_C/(2\pi)$=\textcolor{blue}{3.5~MHz}, 6.98\textcolor{blue}{~MHz} and $12.74$~MHz respectively.
%$1.95696(ea_0)$, $0.01233968(ea_0)$ and $1574.69135769(ea_0)$ respectively

The interaction between atom  and electromagnetic fields is described by writing the Liouville-von Neumann master equation for the density matrix $\rho$ which is expressed as
\begin{eqnarray}
\label{Linb}
&\dot{\rho}=-\frac{i}{\hbar}[\textrm{H}, \rho]-\frac{1}{2}\{\Gamma,\rho\},
\end{eqnarray}
where, $\Gamma$ the relaxation operator defined as $\bra{i}\Gamma\ket{j}=\Gamma_{i}\delta_{ij} ~(\delta_{ij}=1$, if $i=j$ and 0, if $i\neq{j})$ and $\Gamma_{i}$ is the decay rate of state $\ket{i}$. In our system, the natural radiative decay rate of the first excited state $5P_{3/2}$ is $\Gamma_2=2\pi \times$ 6.07~MHz \cite{SAF11}  and that of the Rydberg states $50D_{5/2}$ and $51P_{3/2}$ are $\Gamma_3=2\pi\times$ 2.4~kHz and $\Gamma_4=2\pi\times$ 1.8~kHz, respectively \cite{BET09}. 
It also includes a dephasing rate of atomic coherence defined by $\gamma_{dep}$ to account for the transit time broadening effects because of finite beam sizes and also collisions \cite{meyer2018digital}. The equations of motion for the density matrix is obtained by substituting Eq. \ref{Hami} into Eq. \ref{Linb} resulting in the following time dependent equations:\small
\begin{equation}
\begin{aligned}
\dot{\rho}_{11}&=-i \frac{\Omega _{P}}{2} \rho_{21}+i \frac{\Omega _{P}^*}{2} \rho_{12}+\Gamma _{21} \rho_{22}+\Gamma _{41} \rho_{44},\\
\dot{\rho}_{12}&=-i \frac{\Omega _{P}}{2} (\rho_{22}-\rho_{11})+i \frac{\Omega _{C}^*}{2} \rho_{13}-\gamma_{12}\rho_{12},\\
\dot{\rho}_{13}&=-i \frac{\Omega _{P}}{2} \rho_{23}+i  \frac{\Omega _{C}}{2}\rho_{12}+i \frac{(\Omega _{LO}^*+\Omega_{SIG}^*e^{-i \Delta  t}) }{2} \rho_{14}\\
&\quad-\gamma_{13}\rho_{13},\\
\dot{\rho}_{14}&=-i \frac{\Omega _{P}}{2} \rho_{24}+i \frac{(\Omega _{LO}+\Omega_{SIG}e^{i \Delta  t}) }{2} \rho_{13}-\gamma_{14}\rho_{14},\\
\dot{\rho}_{22}&=-i \frac{\Omega _{P}^*}{2} \rho_{12}-i  \frac{\Omega _{C}}{2} \rho_{32}+i \frac{\Omega _{P}}{2} \rho_{21}+i \frac{\Omega _{C}^*}{2} \rho_{23}-\Gamma_{21}\rho_{22} \\
&\quad+\Gamma_{32}\rho_{33},\\
\dot{\rho}_{23}&=-i \frac{\Omega _{P}^*}{2} \rho_{13}-i  \frac{\Omega _{C}}{2} (\rho_{33}-\rho_{22})+i \frac{(\Omega _{LO}^*+\Omega_{SIG}^*e^{-i \Delta  t}) }{2} \rho_{24} \\
&\quad-\gamma_{23}\rho_{23},\\
\dot{\rho}_{24}&=-i \frac{\Omega _{P}^*}{2} \rho_{14}-i  \frac{\Omega _{C}}{2} \rho_{34}+i \frac{(\Omega _{LO}+\Omega_{SIG}e^{i \Delta  t}) }{2} \rho_{23}-\gamma_{24}\rho_{24},\\
\dot{\rho}_{33}&=-i \frac{\Omega _{C}^*}{2} \rho_{23}-i \frac{(\Omega _{LO}+\Omega_{SIG}e^{i \Delta  t}) }{2} \rho_{43}+i  \frac{\Omega _{C}}{2} \rho_{32}\\
&\quad+i \frac{(\Omega _{LO}^*+\Omega_{SIG}^*e^{-i \Delta  t}) }{2} \rho_{34}-(\Gamma_{32}+\Gamma_{34})\rho_{33},\\
\dot{\rho}_{34}&=-i \frac{\Omega _{C}^*}{2} \rho_{24}+i \frac{(\Omega _{LO}+\Omega_{SIG}e^{i \Delta  t}) }{2}( \rho_{33}-\rho_{44})-\gamma_{34}\rho_{34},\\
\dot{\rho}_{44}&=i \frac{(\Omega _{LO}^*+\Omega_{SIG}^*e^{i \Delta  t}) }{2} \rho_{34}+i \frac{(\Omega _{LO}+\Omega_{SIG}e^{-i \Delta  t}) }{2} \rho_{43}\\
&\quad-\Gamma_{41}\rho_{44},\\
\end{aligned}
\label{den1}
\end{equation}
where $ \gamma_{12}=i\Delta_{P}+\gamma^{dec}_{12}+\gamma_{dep}$, $ \gamma_{13}=i(\Delta_{P}+\Delta_{C})+\gamma^{dec}_{13}+\gamma_{dep}$, $ \gamma_{14}=i(\Delta_{P}+\Delta_{C}-\Delta_{LO})+\gamma^{dec}_{14}+\gamma_{dep}$, $ \gamma_{23}=-i\Delta_{C}+\gamma^{dec}_{32}+\gamma_{dep}$, $ \gamma_{24}=i(\Delta_{C}-\Delta_{LO})+\gamma^{dec}_{24}+\gamma_{dep}$, $ \gamma_{34}=i(-\Delta_{LO})+\gamma^{dec}_{34}+\gamma_{dep}$ and $ \gamma_{ij}^{dec}=\frac{1}{2}(\Gamma_{i}+\Gamma_{j}) $ is the decoherence rate between level $ \ket{i} $ and $ \ket{j} $.

 The time dependent equations can be solved using the Floquet expansion where we expand $\rho$ into harmonics of frequency $\Delta$ and is given as
\begin{eqnarray}
\label{Floq}
\rho_{ij}(t)=&\sum_{n=-\infty}^{\infty}\rho^{(n)}_{ij}(t)e^{in\Delta{t}}, 
\end{eqnarray}
where $\rho^{(n)}_{ij}(t)$ are the $n^{th}$ order harmonic amplitudes. Substituting Eq. \ref{Floq} in Eq. \ref{den1} and comparing coefficients with same power in $e^{in\delta{t}}$ removes the time dependence, yielding the following set of steady-state equations for the slowly varying harmonic amplitudes:\small
\begin{equation}
\begin{aligned}
\dot{\rho}_{11}^{(n)}&=-in\delta \rho_{11}^{(n)}-i \frac{\Omega _{P}}{2} \rho_{21}^{(n)}+i \frac{\Omega _{P}^*}{2} \rho_{12}^{(n)}+\Gamma _{21} \rho_{22}^{(n)}+\Gamma _{41} \rho_{44}^{(n)},\\
\dot{\rho}_{12}^{(n)}&=-in\delta \rho_{12}^{(n)}-i \frac{\Omega _{P}}{2} (\rho_{22}^{(n)}-\rho_{11}^{(n)})+i \frac{\Omega _{C}^*}{2} \rho_{13}^{(n)}-\gamma_{12}\rho_{12}^{(n)},\\
\dot{\rho}_{13}^{(n)}&=-in\delta \rho_{13}^{(n)}-i \frac{\Omega _{P}}{2} \rho_{23}^{(n)}+i  \frac{\Omega _{C}}{2}\rho_{12}^{(n)}+i \frac{\Omega _{LO}}{2} \rho_{14}^{(n)}+i\frac{\Omega_{SIG}}{2} \rho_{14}^{(n+1)} \\
&\quad-\gamma_{13}\rho_{13}^{(n)},\\
\dot{\rho}_{14}^{(n)}&=-in\delta \rho_{14}^{(n)}-i \frac{\Omega _{P}}{2} \rho_{24}^{(n)}+i \frac{\Omega _{LO}}{2} \rho_{13}^{(n)}+\frac{\Omega_{SIG}}{2} \rho_{13}^{(n-1)}-\gamma_{14}\rho_{14}^{(n)},\\
\dot{\rho}_{22}^{(n)}&=-in\delta \rho_{22}^{(n)}-i \frac{\Omega _{P}^*}{2} \rho_{12}^{(n)}-i  \frac{\Omega _{C}}{2} \rho_{32}^{(n)}+i \frac{\Omega _{P}}{2} \rho_{21}^{(n)}+i \frac{\Omega _{C}^*}{2} \rho_{23}^{(n)}\\
&\quad-\Gamma_{21}\rho_{22}^{(n)}+\Gamma_{32}\rho_{33}^{(n)},\\
\dot{\rho}_{23}^{(n)}&=-in\delta \rho_{23}^{(n)}-i \frac{\Omega _{P}^*}{2} \rho_{13}^{(n)}-i  \frac{\Omega _{C}}{2} (\rho_{33}^{(n)}-\rho_{22}^{(n)})+i \frac{\Omega _{LO}^*}{2} \rho_{24}^{(n)} \\
&\quad+i\frac{\Omega_{SIG}^*}{2} \rho_{24}^{(n+1)}-\gamma_{23}\rho_{23}^{(n)},\\
\dot{\rho}_{24}^{(n)}&=-in\delta \rho_{24}^{(n)}-i \frac{\Omega _{P}^*}{2} \rho_{14}^{(n)}-i  \frac{\Omega _{C}}{2} \rho_{34}^{(n)}+i \frac{\Omega _{LO}}{2} \rho_{23}^{(n)}+i\frac{\Omega_{SIG}}{2} \rho_{23}^{(n-1)}\ \\
&\quad-\gamma_{24}\rho_{24}^{(n)},\\
\dot{\rho}_{33}^{(n)}&=-in\delta \rho_{33}^{(n)}-i \frac{\Omega _{C}^*}{2} \rho_{23}^{(n)}-i \frac{\Omega _{LO}}{2} \rho_{43}^{(n)}-i\frac{\Omega_{SIG}}{2} \rho_{43}^{(n-1)}+i  \frac{\Omega _{C}}{2} \rho_{32}^{(n)} \\
&\quad+i \frac{\Omega _{LO}^*}{2} \rho_{34}^{(n)}+i\frac{\Omega_{SIG}^*}{2} \rho_{34}^{(n+1)}-(\Gamma_{32}-\Gamma_{34})\rho_{33}^{(n)},\\
\dot{\rho}_{34}^{(n)}&=-in\delta \rho_{34}^{(n)}-i \frac{\Omega _{C}^*}{2} \rho_{24}^{(n)}+i \frac{\Omega _{LO}}{2}( \rho_{33}^{(n)}-\rho_{44}^{(n)})\\
&\quad+i\frac{\Omega_{SIG}}{2} (\rho_{33}^{(n-1)}-\rho_{44}^{(n-1)})-\gamma_{34}\rho_{34}^{(n)},\\
\dot{\rho}_{44}^{(n)}&=-in\delta \rho_{44}^{(n)}+i \frac{\Omega _{LO}^*}{2} \rho_{34}^{(n)}+i\frac{\Omega_{SIG}^*}{2} \rho_{34}^{(n-1)}+i \frac{\Omega _{LO}}{2} \rho_{43}^{(n)} \\
&\quad+i\frac{\Omega_{SIG}}{2} \rho_{43}^{(n+1)}-\Gamma_{41}\rho_{44}^{(n)},\\
\end{aligned}
\label{den2}
\end{equation}

It is difficult to derive an analytical solution of the various harmonic components of the density matrix without the weak probe approximation but the equations can be solved numerically and we have truncated the order at n=1 because for small RF signal field, the amplitude of the higher harmonics is very small. The $\text{Im}(\rho^{(0)}_{12})$ and $\text{Im}(\rho^{(\pm1)}_{12})$ gives the zeroth order harmonics and first order harmonics of the probe absorption. The first order  harmonics, $\text{Im}(\rho^{(-1)}_{12})$ and $\text{Im}(\rho^{(+1)}_{12})$ describes the two generated negative and positive sidebands.

 To account for the thermal motion of atoms, doppler averaging of $\text{Im}(\rho^{(\pm1)}_{12})$ is done numerically at the room temperature (T=300 K).  This is done by integrating $\text{Im}(\rho^{(\pm1)}_{12})$ over the velocities weighted by the Maxwell Boltzman velocity distribution function as 
\begin{equation}
\langle \text{Im}( \rho^{\pm}_{12} )\rangle = \frac{1}{\sqrt{\pi} v_p} \int_{-3v_p}^{+3v_p}\text{Im}(\rho_{12}^{\pm}(v) )e^{-v^2/v_p^2} dv,
\label{eq:Dop}
\end{equation}
where  $v_p=\sqrt{2k_BT/m}$ is the most probable velocity of the atom, $k_B$ is Boltzman constant and $m$ is atomic mass of $^{85}$Rb. %The choice of limits for numerical integration is set to $\pm3v_P$ which corresponds to a velocity distribution amplitude of $1/e^9$. 
For a moving atoms the detunings are modifed by replacing $\Delta_P$ with $\Delta_P + k_{P}v$ and $\Delta_C$ with $\Delta_C - k_{C}v$ for the counter-propagating configuration, while the Doppler shift for the MW fields are ignored. Here,  $k_P$ and $k_C$ are the wave vectors of probe and coupling lasers and v is the velocity of the atom. We use the normalised form of Eq. \ref{eq:Dop} which is 
$\langle\text{Im}(\rho^-_{12})\rangle$
% $<\text{Im}(\rho^-_{12})>$
and 
$\langle\text{Im}(\rho^+_{12})\rangle$
% $<\text{Im}(\rho^+_{12})>$
to simulate our theoretical bandwidth curves for the negative and positive sidebands, respectively.

\section{Results and Discussions}\label{sec:res}

\begin{figure}[h]
    \centering
    \includegraphics[width=1.0\columnwidth]{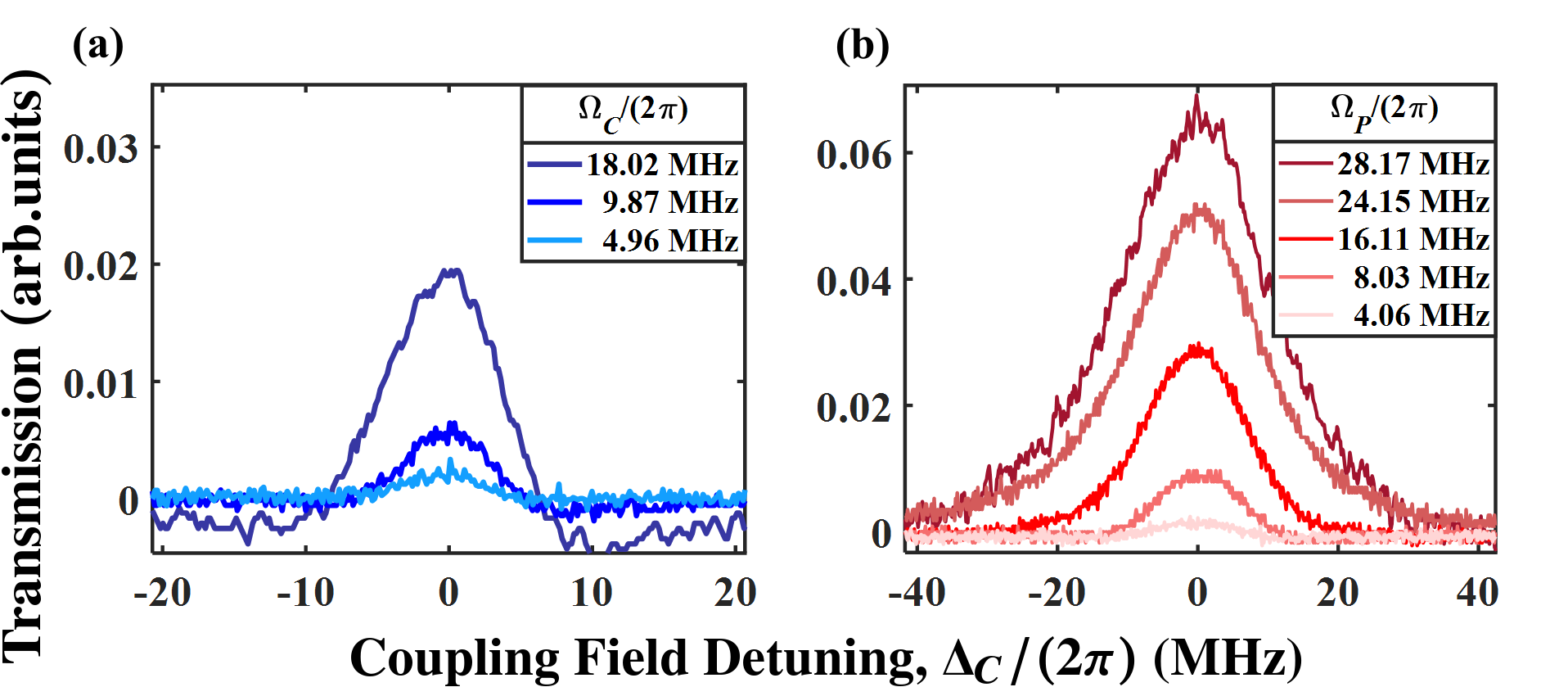}
\caption{Probe transmission as a function of coupling field detuning, $\Delta_C/(2\pi)$ in ~MHz with the RF fields off. The different plots represents: (a) different coupling Rabi frequencies at fixed probe Rabi frequency, $\Omega_P/2\pi$ = 1.66~MHz and (b) different probe Rabi frequencies at fixed coupling Rabi frequency $\Omega_C/2\pi$ = 18.02~MHz, as labelled in the legend.}
    \label{fig:EIT_CP}
\end{figure}

We initially studied the effects of the coupling Rabi frequency and then move on to study the effects of the probe Rabi frequency on the  bandwidth of our system. We also compared our experimental bandwidth results to our theoretical model. Figure \ref{fig:EIT_CP} (a) shows the EIT spectra for the three sets of coupling Rabi frequencies, $\Omega_{C}/2\pi$ = 4.96~MHz, 9.87~MHz and 18.02~MHz with RF fields off and keeping the probe Rabi frequency fixed at $\Omega_P/(2\pi)=$ 1.66~MHz and Figure \ref{fig:EIT_CP} (b) shows the EIT spectra for the five sets of probe Rabi frequencies, $\Omega_{P}/2\pi$ = 4.06~MHz, 8.03~MHz, 16.11~MHz, 24.15~MHz and 28.17~MHz with RF fields off and keeping the coupling Rabi frequency fixed at $\Omega_C/(2\pi)=$ 18.02~MHz. The full-width at half-maximum (FWHM) of the EIT spectra obtained from the fits for figure \ref{fig:EIT_CP} (a) are $\sim$5.24~MHz, 9.86~MHz and 11.47~MHz, and for figure \ref{fig:EIT_CP} (b) are 12.19~MHz, 15.72~MHz, 19.70~MHz, 24.10~MHz and 28.42~MHz, respectively.
%%%%%%%%%%%%%%%%%%%%Figure2%%%%%%%%%%%%%%%%%%%%%%%%%%%
%%%%%%%%%%%%%%%%%%%%Figure3%%%%%%%%%%%%%%%%%%%%%%%%%%%%
\begin{figure}[h]
    \centering
    \includegraphics[width=0.85\columnwidth]{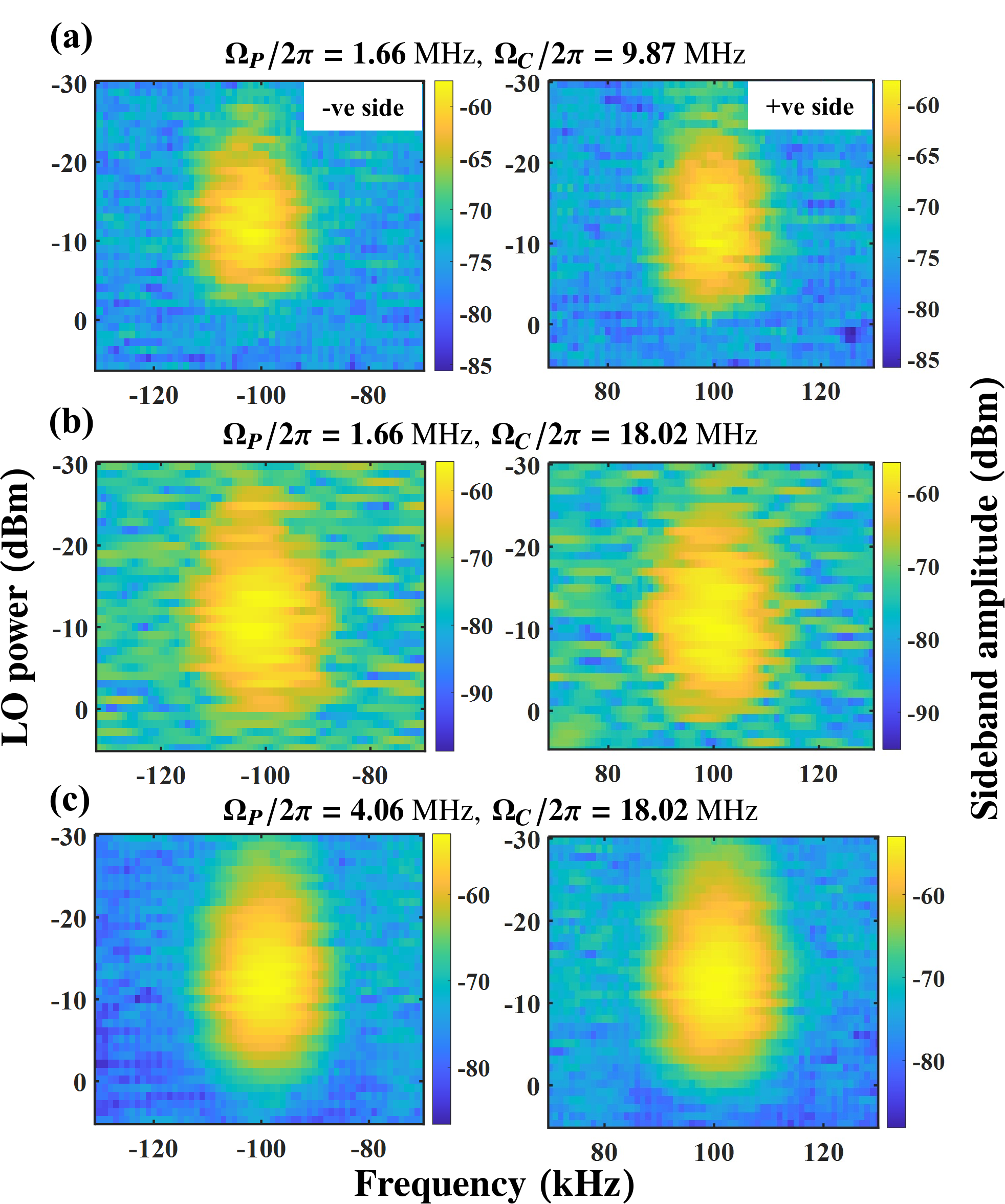}
    \caption{Color map of the sideband amplitude versus the RF LO power sweep in dBm keeping the RF signal power fixed at $-24$~dBm and detuning at $\Delta_{SIG}=100$~kHz. The figures on the left and right shows the negative and positive generated sidebands at frequency $\pm100$ kHz for the different combination of probe and coupling Rabi frequencies: (a) $\Omega_P/(2\pi)$ = 1.66~MHz, $\Omega_C/(2\pi)$ = 9.87~MHz, (b) $\Omega_P/(2\pi)$ = 1.66~MHz, $\Omega_C/(2\pi)$ = 18.02~MHz and (c) $\Omega_P/(2\pi)$ = 4.06~MHz, $\Omega_C/(2\pi)$ = 18.02~MHz.}
    \label{fig:LO_OPT1}
\end{figure}
%%%%%%%%%%%%%%%%%%%%%Figure4%%%%%%%%%%%%%%%%%%%%%%%%%%%%%
\begin{figure}[h]
    \centering
    \includegraphics[width=0.85\columnwidth]{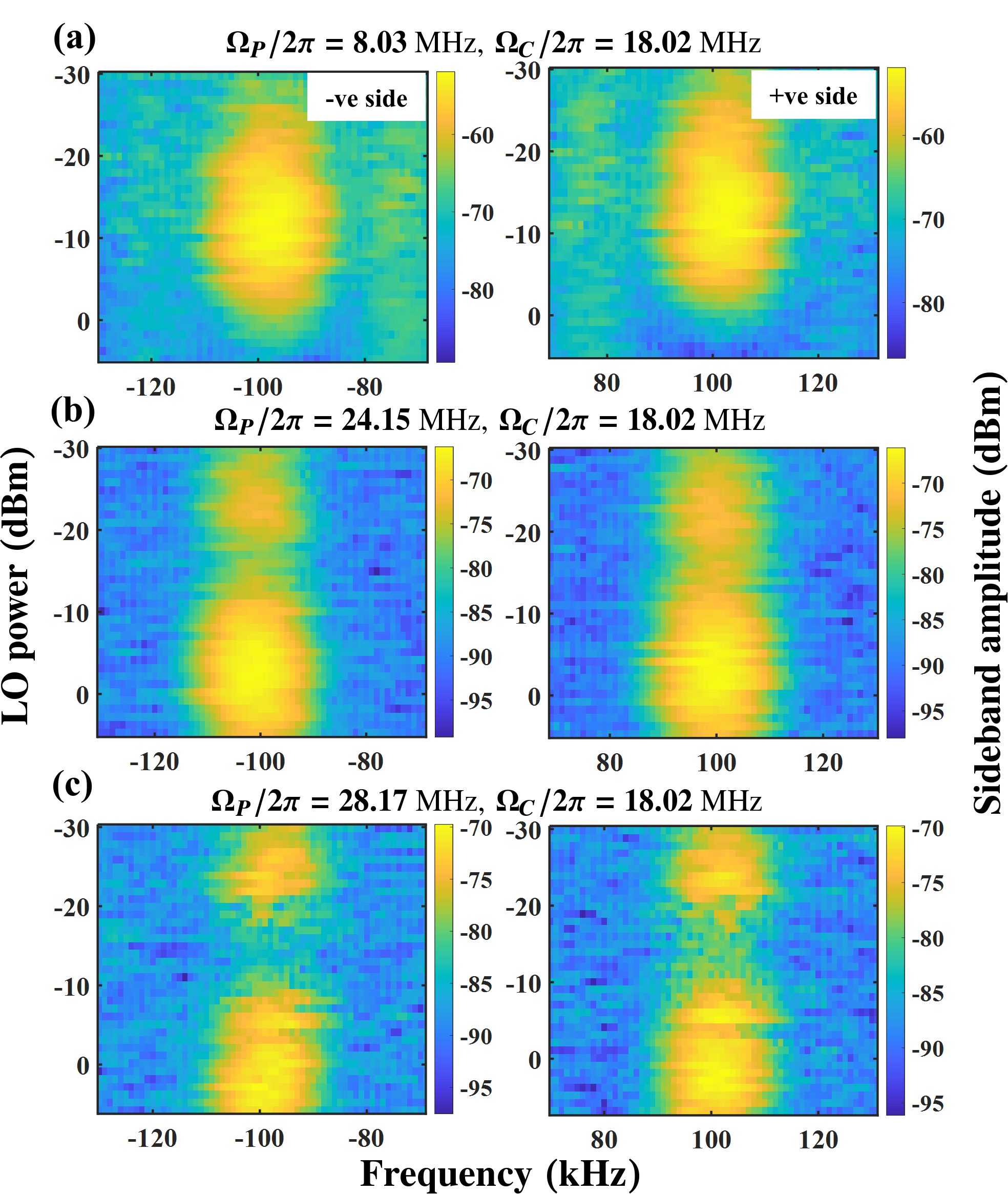}
    \caption{Color map of the sideband amplitude versus the RF LO power sweep in dBm keeping the RF signal power fixed at $-24$~dBm and detuning at $\Delta_{SIG}=100$~kHz. The figures on the left and right shows the negative and positive generated sidebands at frequency $\pm100$ kHz for the different combination of probe and coupling Rabi frequencies: (a) $\Omega_P/(2\pi)$ = 8.03~MHz, $\Omega_C/(2\pi)$=18.02~MHz, (b) $\Omega_P/(2\pi)$ = 24.15~MHz, $\Omega_C/(2\pi)$ = 18.02~MHz and (c) $\Omega_P/(2\pi)$ = 28.17~MHz, $\Omega_C/(2\pi)$ = 18.02~MHz.}
    \label{fig:LO_OPT2}
\end{figure}
%%%%%%%%%%%%%%%%%%%%%%Figure5%%%%%%%%%%%%%%%%%%%%%%%%%%%%%%
\begin{figure}
    \centering
    \includegraphics[width=1.04\columnwidth]{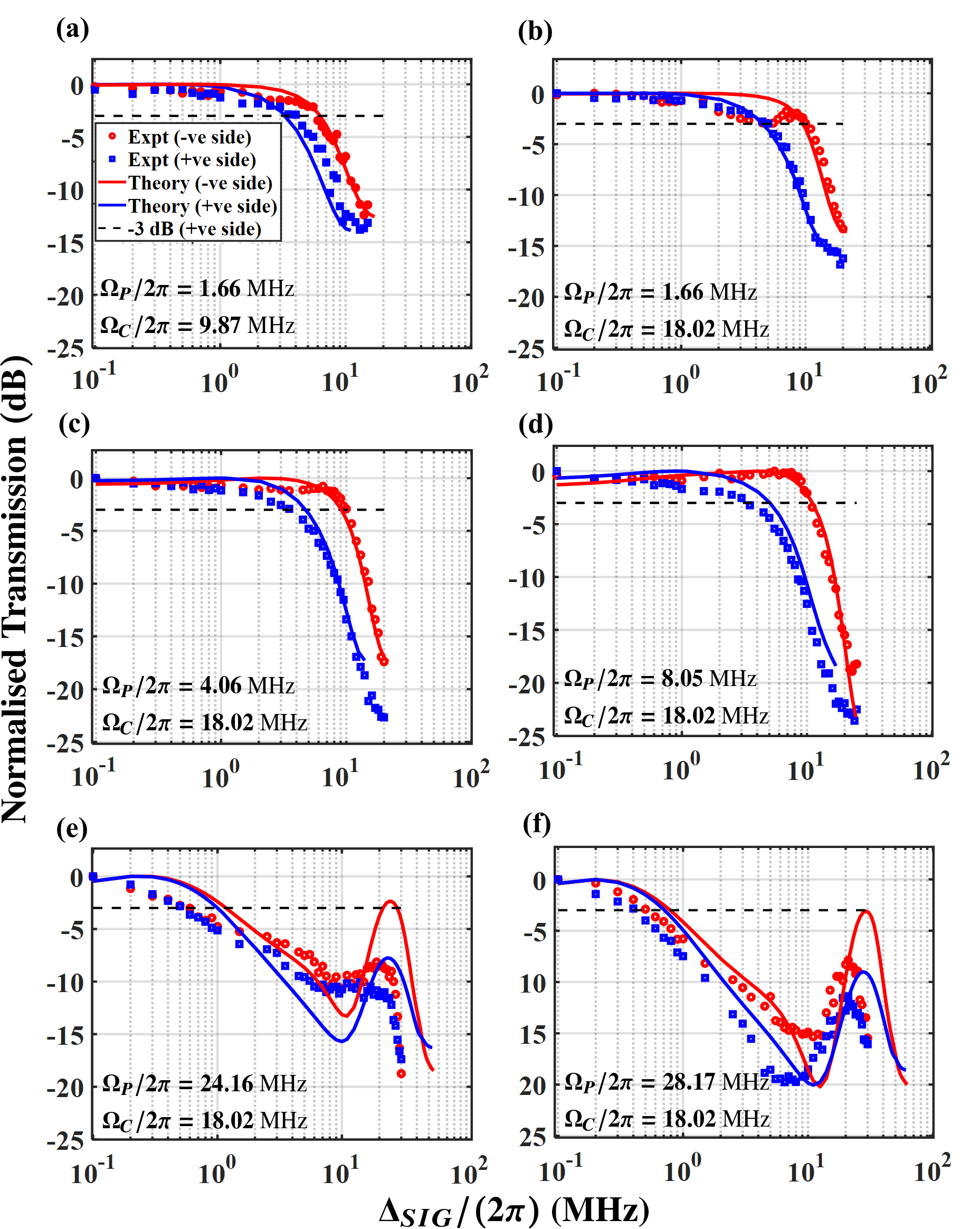}
    \caption{Normalised transmission in dB as a function of $\Delta_{SIG}/(2\pi)$ in ~MHz for the different combination of probe and coupling Rabi frequencies. (a) $\Omega_P/(2\pi)$ = 1.66~MHz, $\Omega_C/(2\pi)$ = 9.87~MHz, (b) $\Omega_P/(2\pi)$ = 1.66~MHz, $\Omega_C/(2\pi)$ = 18.02~MHz and (c) $\Omega_P/(2\pi)$ = 4.06~MHz, $\Omega_C/(2\pi)$ = 18.02~MHz, (d) $\Omega_P/(2\pi)$ = 8.05~MHz, $\Omega_C/(2\pi)$ = 18.02~MHz, (e) $\Omega_P/(2\pi)$ = 24.16~MHz, $\Omega_C/(2\pi)$ = 18.02~MHz and (f) $\Omega_P/(2\pi)$ = 28.17~MHz, $\Omega_C/(2\pi)$= 18.02~MHz.  The dotted and solid lines represent experimental and theoretical data, respectively. The colour-coded data, with red and blue represents the negative and positive sidebands, respectively, also includes a dashed line marking the -3dB line.}
    \label{fig:BW_comp}
\end{figure}

%%%%%%%%%%%%%%%%%%%%%%%%%%%%%%%%%%%%%%%%%%%%%%%%%%%%%%

%%%%%%%%%%%%%%%%%%%%%%Figure6%%%%%%%%%%%%%%%%%%%%%%%%%%%%%%
\begin{figure}
    \centering
    \includegraphics[width=1.05\columnwidth]{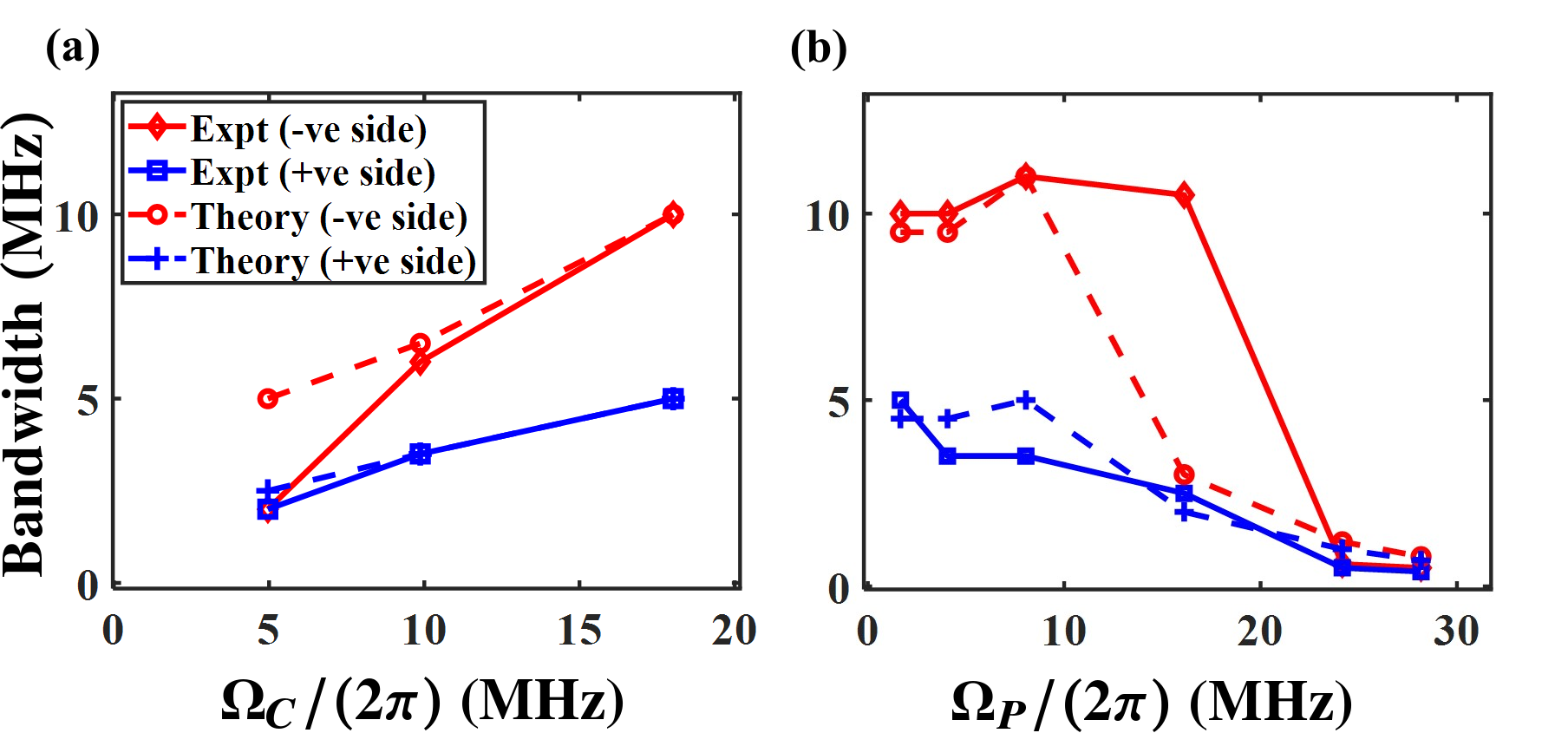}
    \caption{ Comparison between experimental and theoretical bandwidths as a function of (a) coupling Rabi frequencies at fixed probe Rabi frequency, $\Omega_P/(2\pi)$= 1.66~MHz and (b) probe Rabi frequencies at fixed coupling Rabi frequency, $\Omega_C/(2\pi)$= 18.02~MHz.}
    \label{fig:BW_rabi}
\end{figure}

%%%%%%%%%%%%%%%%%%%%%%%%%%%%%%%%%%%%%%%%%%%%%%%%%%%%%%
 
We first optimized the RF LO field power for the different coupling Rabi frequencies keeping the probe Rabi frequency fixed. Figure \ref{fig:LO_OPT1} (a) and (b) presents a color map of the negative (left) and positive (right) sideband amplitude as a function of LO power. The RF signal field power is kept fixed at -24 dBm and frequency detuning $\Delta_{SIG}= 100$ kHz, while the probe Rabi frequency is kept fixed at $\Omega_P/(2\pi)=$ 1.66~MHz for the coupling Rabi frequencies $\Omega_{C}/2\pi$ = 8.03~MHz and 18.02~MHz. We see that the RF LO field has an optimized power value, where the sideband amplitudes reaches a maximum indicating where it most sensitive to the RF signal field. Also, the optimized LO field power increases with increase in the coupling Rabi frequencies. Using these plots, we fix the power of the RF LO field at its optimized point for our bandwidth measurements. The LO optimization occurs at -13 and -10 dBm which corresponds to RF LO Rabi frequencies, $\Omega_{LO}/(2\pi)=$ 3.53 and 5.05 ~MHz, respectively. These values were determined by calibrating the RF LO power to the resulting Rabi frequencies, based on the measured spectral splitting of the AT peaks as the coupling laser is scanned over resonance.  %\textcolor{red}{enter the converted dbm to MHz LO Rabis}

Similarly, we optimized the RF LO field power for different probe Rabi frequencies keeping coupling Rabi frequency fixed.  Figure \ref{fig:LO_OPT1} (c), \ref{fig:LO_OPT2} (a), (b) and (c) shows a color map of the negative and positive sideband amplitudes as a function of LO power keeping coupling Rabi frequency fixed at $\Omega_C/(2\pi) =$ 18.02~MHz for different probe Rabi frequencies at $\Omega_P/(2\pi) =$ 4.06~MHz,  8.03~MHz, 24.15~MHz and 28.17~MHz. We see that the RF LO field power also has an optimize point which increases with increase in the probe Rabi frequency. At high probe Rabi frequencies, specifically for  $\Omega_P/(2\pi) =$ 24.15 and 28.17~MHz a second maxima appears when $\Omega_{LO}\approx \Omega_{SIG}$ because in this region as the LO power increases, the EIT peak initially rises to a maximum and then starts to split. We choose  the optimized RF LO field when $\Omega_{LO}> \Omega_{SIG}$ for our bandwidth measurements as it is most sensitive for this case. The LO optimization for this case occurs at -10, -9, -1 and 0~dBm which corresponds to RF LO Rabi frequencies, $\Omega_{LO}/(2\pi)=$ 5.05, 5.61, 14.09 and 15.81 ~MHz, respectively.  

%When detuning the RF signal relative to the LO field frequency, we have two options: red detuning, $\Delta_{SIG}<0$ and blue detuning, $\Delta_{SIG}>0$. Both methods yield comparable results. Therefore, we only discuss results from the $\Delta_{SIG}$$<$ 0 case. 
We then measured the bandwidth of our system by plotting the experimental normalised transmission in dB as a function of $\Delta_{SIG}/(2\pi)$ in MHz for the different combination of probe and coupling Rabi frequencies, overlaid with the theoretical normalized transmission, $\langle\text{Im}(\rho^\pm_{12})\rangle$ and  is shown in Figure \ref{fig:BW_comp}. The dotted and solid lines represent experimental and theoretical data, respectively. The red and blue represents the negative and positive sidebands, respectively. A dashed line marks the -3dB line from where we extract the roll-off point of the bandwidth of our system. 

We observe that the bandwidth increases with the increase in coupling Rabi frequency while keeping the probe Rabi frequency small which is a requirement of the system for high bandwidth as previous studies have shown that higher coupling power significantly enhances bandwidth \cite{yang2023high,Hu2023,AlyFCIB24}. We also see that the bandwidth of the negative sideband is larger than that of the positive sideband and is shown in Figure \ref{fig:BW_comp} (a) and (b). The experimental (and theoretical) bandwidths extracted from these plots are shown in Figure~\ref{fig:BW_rabi} (a): 6 (6.5) and 3.5 (3.5)~MHz for negative and positive sidebands at coupling Rabi frequency of $\Omega_C/(2\pi) =$ 9.87~MHz and  10 (9.5) and 5 (4.5)~MHz for negative and positive sidebands at coupling Rabi frequency of $\Omega_C/(2\pi)$ = 18.02~MHz keeping probe Rabi frequency fixed at $\Omega_P/(2\pi) =$ 1.66~MHz. 

But as we increase the probe Rabi frequency keeping the coupling Rabi frequency fixed, the bandwidth specifically at $\Omega_{P}$ = 24.15~MHz and 28.17~MHz, decreases and at the same time we see a high frequency response peak as shown in Figure \ref{fig:BW_comp} (e) and (f). The decrease in bandwidth with the increase in probe Rabi frequency can be attributed to the fact that a high probe Rabi frequency leads to a slower instantaneous response \cite{bohaichuk2022origins}. The experimental (and theoretical) bandwidths extracted from these plots are shown in Figure~\ref{fig:BW_rabi} (b): 10 (9.5) and 3.5 (4.5)~MHz for the negative and positive sidebands at probe Rabi frequency of $\Omega_P/(2\pi) =$ 4.06~MHz,  11 (11)~MHz and 3.5 (5)~MHz for negative and positive sidebands at probe Rabi frequencies of $\Omega_P/(2\pi) =$ 8.03~MHz, 0.6 (1.2)~MHz  and 0.5 (1)~MHz for negative and positive sidebands at probe Rabi frequency of $\Omega_P/(2\pi) =$  24.15~MHz,  0.5 (0.8)~MHz for negative and 0.4 (0.7) MHz for positive sidebands at probe Rabi frequencies of $\Omega_P/(2\pi) =$ 28.17~MHz.  To ensure consistency between theoretical and experimental results, we set the dephasing rate in the theoretical model at $\gamma_{dep}$ = 5~MHz for all cases. 
At high probe Rabi frequency, the theory suggests a slightly higher response bandwidth than what we observe in experiments, this is possibly due to phase noise in the lasers. 
We also note that a large difference between the theoretical and experimental bandwidths is seen in Figure~\ref{fig:BW_rabi} (b) at $\Omega_P/(2\pi)=16.11$~MHz. %\textcolor{red}{(what's the exact frequency here?)}.
We attribute this to the emergence of the signal increase at high values of $\Delta_{SIG}$ in the regime where $\Omega_P>\Omega_C$, which is poorly described by our theory in the crossover regime where $\Omega_P \approx \Omega_C$.
%\textcolor{red}{Conclusion could use some restructuring/streamlining. Is there any discussion of this signal increase with $\Delta_{SIG}$ in the regime where $\Omega_P>\Omega_C$? This is distinct feature and should at least be addressed.}

\section{Conclusion}\label{sec:conc}

We have investigated the bandwidth of a Rydberg atom-based receiver in an LO-based Rydberg mixer scheme through optical heterodyning of the probe laser.
With this method, we were able to differentiate between the negative and the positive generated sidebands attributed to distinct six-wave mixing processes and to confirm previous theoretical predictions that the negative sideband exhibits a higher bandwidth compared to the positive sideband at higher beat note frequencies which we find to be in good agreement with our theoretical model.
We see that the bandwidth increases with the coupling Rabi frequency, while it decreases with the probe Rabi frequency followed by a high frequency response peak. 
By optimizing the field parameters, we achieved a maximum experimental (and theoretical) bandwidth of the Rydberg receiver about 11 (11)~MHz and 3.5 (5)~MHz for the negative and positive sideband from the -3dB roll-off point. 
Our findings provide valuable insights into bandwidth-limiting atomic features, which can potentially advance future atomic Rydberg receiving technology. 

\section*{Acknowledgments}
This work was supported by DARPA under the Quantum Apertures program. The views, opinions and/or findings expressed are those of the authors and should not be interpreted as representing the official views or policies of the Department of Defense or the U.S. Government. A contribution of the U.S. government, this work is not subject to copyright in the United States.

\subsection*{Conflict of Interest}
\vspace{-3mm}
The authors have no conflicts to disclose.
\vspace{-3mm}
\subsection*{Data Availability Statement}
\vspace{-3mm}
All data presented in this review will be made available to the reader upon reasonable request to the authors.

\bibliography{cites}

\begin{thebibliography}{20}%
\makeatletter
\providecommand \@ifxundefined [1]{%
 \@ifx{#1\undefined}
}%
\providecommand \@ifnum [1]{%
 \ifnum #1\expandafter \@firstoftwo
 \else \expandafter \@secondoftwo
 \fi
}%
\providecommand \@ifx [1]{%
 \ifx #1\expandafter \@firstoftwo
 \else \expandafter \@secondoftwo
 \fi
}%
\providecommand \natexlab [1]{#1}%
\providecommand \enquote  [1]{``#1''}%
\providecommand \bibnamefont  [1]{#1}%
\providecommand \bibfnamefont [1]{#1}%
\providecommand \citenamefont [1]{#1}%
\providecommand \href@noop [0]{\@secondoftwo}%
\providecommand \href [0]{\begingroup \@sanitize@url \@href}%
\providecommand \@href[1]{\@@startlink{#1}\@@href}%
\providecommand \@@href[1]{\endgroup#1\@@endlink}%
\providecommand \@sanitize@url [0]{\catcode `\\12\catcode `\$12\catcode
  `\&12\catcode `\#12\catcode `\^12\catcode `\_12\catcode `\%12\relax}%
\providecommand \@@startlink[1]{}%
\providecommand \@@endlink[0]{}%
\providecommand \url  [0]{\begingroup\@sanitize@url \@url }%
\providecommand \@url [1]{\endgroup\@href {#1}{\urlprefix }}%
\providecommand \urlprefix  [0]{URL }%
\providecommand \Eprint [0]{\href }%
\providecommand \doibase [0]{https://doi.org/}%
\providecommand \selectlanguage [0]{\@gobble}%
\providecommand \bibinfo  [0]{\@secondoftwo}%
\providecommand \bibfield  [0]{\@secondoftwo}%
\providecommand \translation [1]{[#1]}%
\providecommand \BibitemOpen [0]{}%
\providecommand \bibitemStop [0]{}%
\providecommand \bibitemNoStop [0]{.\EOS\space}%
\providecommand \EOS [0]{\spacefactor3000\relax}%
\providecommand \BibitemShut  [1]{\csname bibitem#1\endcsname}%
\let\auto@bib@innerbib\@empty
%</preamble>
\bibitem [{\citenamefont {Sedlacek}\ \emph {et~al.}(2012)\citenamefont
  {Sedlacek}, \citenamefont {Schwettmann}, \citenamefont {K{\"u}bler},
  \citenamefont {L{\"o}w}, \citenamefont {Pfau},\ and\ \citenamefont
  {Shaffer}}]{sedlacek2012microwave}%
  \BibitemOpen
  \bibfield  {author} {\bibinfo {author} {\bibfnamefont {J.~A.}\ \bibnamefont
  {Sedlacek}}, \bibinfo {author} {\bibfnamefont {A.}~\bibnamefont
  {Schwettmann}}, \bibinfo {author} {\bibfnamefont {H.}~\bibnamefont
  {K{\"u}bler}}, \bibinfo {author} {\bibfnamefont {R.}~\bibnamefont {L{\"o}w}},
  \bibinfo {author} {\bibfnamefont {T.}~\bibnamefont {Pfau}},\ and\ \bibinfo
  {author} {\bibfnamefont {J.~P.}\ \bibnamefont {Shaffer}},\ }\bibfield
  {title} {\bibinfo {title} {Microwave electrometry with rydberg atoms in a
  vapour cell using bright atomic resonances},\ }\href@noop {} {\bibfield
  {journal} {\bibinfo  {journal} {Nature physics}\ }\textbf {\bibinfo {volume}
  {8}},\ \bibinfo {pages} {819} (\bibinfo {year} {2012})}\BibitemShut {NoStop}%
\bibitem [{\citenamefont {Holloway}\ \emph {et~al.}(2014)\citenamefont
  {Holloway}, \citenamefont {Gordon}, \citenamefont {Jefferts}, \citenamefont
  {Schwarzkopf}, \citenamefont {Anderson}, \citenamefont {Miller},
  \citenamefont {Thaicharoen},\ and\ \citenamefont {Raithel}}]{6910267}%
  \BibitemOpen
  \bibfield  {author} {\bibinfo {author} {\bibfnamefont {C.~L.}\ \bibnamefont
  {Holloway}}, \bibinfo {author} {\bibfnamefont {J.~A.}\ \bibnamefont
  {Gordon}}, \bibinfo {author} {\bibfnamefont {S.}~\bibnamefont {Jefferts}},
  \bibinfo {author} {\bibfnamefont {A.}~\bibnamefont {Schwarzkopf}}, \bibinfo
  {author} {\bibfnamefont {D.~A.}\ \bibnamefont {Anderson}}, \bibinfo {author}
  {\bibfnamefont {S.~A.}\ \bibnamefont {Miller}}, \bibinfo {author}
  {\bibfnamefont {N.}~\bibnamefont {Thaicharoen}},\ and\ \bibinfo {author}
  {\bibfnamefont {G.}~\bibnamefont {Raithel}},\ }\bibfield  {title} {\bibinfo
  {title} {Broadband {Rydberg} atom-based electric-field probe for
  si-traceable, self-calibrated measurements},\ }\href
  {https://doi.org/10.1109/TAP.2014.2360208} {\bibfield  {journal} {\bibinfo
  {journal} {IEEE Transactions on Antennas and Propagation}\ }\textbf {\bibinfo
  {volume} {62}},\ \bibinfo {pages} {6169} (\bibinfo {year}
  {2014})}\BibitemShut {NoStop}%
\bibitem [{\citenamefont {Artusio-Glimpse}\ \emph {et~al.}(2022)\citenamefont
  {Artusio-Glimpse}, \citenamefont {Simons}, \citenamefont {Prajapati},\ and\
  \citenamefont {Holloway}}]{9748947}%
  \BibitemOpen
  \bibfield  {author} {\bibinfo {author} {\bibfnamefont {A.}~\bibnamefont
  {Artusio-Glimpse}}, \bibinfo {author} {\bibfnamefont {M.~T.}\ \bibnamefont
  {Simons}}, \bibinfo {author} {\bibfnamefont {N.}~\bibnamefont {Prajapati}},\
  and\ \bibinfo {author} {\bibfnamefont {C.~L.}\ \bibnamefont {Holloway}},\
  }\bibfield  {title} {\bibinfo {title} {Modern rf measurements with hot atoms:
  A technology review of {Rydberg} atom-based radio frequency field sensors},\
  }\href {https://doi.org/10.1109/MMM.2022.3148705} {\bibfield  {journal}
  {\bibinfo  {journal} {IEEE Microwave Magazine}\ }\textbf {\bibinfo {volume}
  {23}},\ \bibinfo {pages} {44} (\bibinfo {year} {2022})}\BibitemShut {NoStop}%
\bibitem [{\citenamefont {Cox}\ \emph {et~al.}(2018{\natexlab{a}})\citenamefont
  {Cox}, \citenamefont {Meyer}, \citenamefont {Fatemi},\ and\ \citenamefont
  {Kunz}}]{cox2018quantum}%
  \BibitemOpen
  \bibfield  {author} {\bibinfo {author} {\bibfnamefont {K.~C.}\ \bibnamefont
  {Cox}}, \bibinfo {author} {\bibfnamefont {D.~H.}\ \bibnamefont {Meyer}},
  \bibinfo {author} {\bibfnamefont {F.~K.}\ \bibnamefont {Fatemi}},\ and\
  \bibinfo {author} {\bibfnamefont {P.~D.}\ \bibnamefont {Kunz}},\ }\bibfield
  {title} {\bibinfo {title} {Quantum-limited atomic receiver in the
  electrically small regime},\ }\href@noop {} {\bibfield  {journal} {\bibinfo
  {journal} {Physical Review Letters}\ }\textbf {\bibinfo {volume} {121}},\
  \bibinfo {pages} {110502} (\bibinfo {year} {2018}{\natexlab{a}})}\BibitemShut
  {NoStop}%
\bibitem [{\citenamefont {Song}\ \emph {et~al.}(2019)\citenamefont {Song},
  \citenamefont {Liu}, \citenamefont {Liu}, \citenamefont {Zhang},
  \citenamefont {Zou}, \citenamefont {Zhang},\ and\ \citenamefont
  {Qu}}]{song1}%
  \BibitemOpen
  \bibfield  {author} {\bibinfo {author} {\bibfnamefont {Z.}~\bibnamefont
  {Song}}, \bibinfo {author} {\bibfnamefont {H.}~\bibnamefont {Liu}}, \bibinfo
  {author} {\bibfnamefont {X.}~\bibnamefont {Liu}}, \bibinfo {author}
  {\bibfnamefont {W.}~\bibnamefont {Zhang}}, \bibinfo {author} {\bibfnamefont
  {H.}~\bibnamefont {Zou}}, \bibinfo {author} {\bibfnamefont {J.}~\bibnamefont
  {Zhang}},\ and\ \bibinfo {author} {\bibfnamefont {J.}~\bibnamefont {Qu}},\
  }\bibfield  {title} {\bibinfo {title} {Rydberg-atom-based digital
  communication using a continuously tunable radio-frequency carrier},\
  }\href@noop {} {\bibfield  {journal} {\bibinfo  {journal} {Optics Express}\
  }\textbf {\bibinfo {volume} {27}},\ \bibinfo {pages} {8848} (\bibinfo {year}
  {2019})}\BibitemShut {NoStop}%
\bibitem [{\citenamefont {Meyer}\ \emph
  {et~al.}(2018{\natexlab{a}})\citenamefont {Meyer}, \citenamefont {Cox},
  \citenamefont {Fatemi},\ and\ \citenamefont {Kunz}}]{meyer1}%
  \BibitemOpen
  \bibfield  {author} {\bibinfo {author} {\bibfnamefont {D.}~\bibnamefont
  {Meyer}}, \bibinfo {author} {\bibfnamefont {K.}~\bibnamefont {Cox}}, \bibinfo
  {author} {\bibfnamefont {F.}~\bibnamefont {Fatemi}},\ and\ \bibinfo {author}
  {\bibfnamefont {P.}~\bibnamefont {Kunz}},\ }\bibfield  {title} {\bibinfo
  {title} {Digital communication with rydberg atoms and amplitude-modulated
  microwave fields},\ }\href@noop {} {\bibfield  {journal} {\bibinfo  {journal}
  {Applied Physics Letters}\ }\textbf {\bibinfo {volume} {112}},\ \bibinfo
  {pages} {211108} (\bibinfo {year} {2018}{\natexlab{a}})}\BibitemShut
  {NoStop}%
\bibitem [{\citenamefont {Cox}\ \emph {et~al.}(2018{\natexlab{b}})\citenamefont
  {Cox}, \citenamefont {Meyer}, \citenamefont {Fatemi},\ and\ \citenamefont
  {Kunz}}]{cox1}%
  \BibitemOpen
  \bibfield  {author} {\bibinfo {author} {\bibfnamefont {K.}~\bibnamefont
  {Cox}}, \bibinfo {author} {\bibfnamefont {D.}~\bibnamefont {Meyer}}, \bibinfo
  {author} {\bibfnamefont {F.}~\bibnamefont {Fatemi}},\ and\ \bibinfo {author}
  {\bibfnamefont {P.}~\bibnamefont {Kunz}},\ }\bibfield  {title} {\bibinfo
  {title} {Quantum-limited atomic receiver in the electrically small regime},\
  }\href@noop {} {\bibfield  {journal} {\bibinfo  {journal} {Phys. Rev. Lett.}\
  }\textbf {\bibinfo {volume} {121}},\ \bibinfo {pages} {110502} (\bibinfo
  {year} {2018}{\natexlab{b}})}\BibitemShut {NoStop}%
\bibitem [{\citenamefont {Holloway}\ \emph {et~al.}(2021)\citenamefont
  {Holloway}, \citenamefont {Simons}, \citenamefont {Haddab}, \citenamefont
  {Gordon}, \citenamefont {Anderson},\ and\ \citenamefont {Raithel}}]{holl4}%
  \BibitemOpen
  \bibfield  {author} {\bibinfo {author} {\bibfnamefont {C.}~\bibnamefont
  {Holloway}}, \bibinfo {author} {\bibfnamefont {M.}~\bibnamefont {Simons}},
  \bibinfo {author} {\bibfnamefont {A.}~\bibnamefont {Haddab}}, \bibinfo
  {author} {\bibfnamefont {J.}~\bibnamefont {Gordon}}, \bibinfo {author}
  {\bibfnamefont {D.}~\bibnamefont {Anderson}},\ and\ \bibinfo {author}
  {\bibfnamefont {G.}~\bibnamefont {Raithel}},\ }\bibfield  {title} {\bibinfo
  {title} {A multiple-band rydberg atom-based receiver: Am/fm stereo
  reception},\ }\href@noop {} {\bibfield  {journal} {\bibinfo  {journal} {IEEE
  Antenna and Propogation Magazine}\ }\textbf {\bibinfo {volume} {63}},\
  \bibinfo {pages} {63} (\bibinfo {year} {2021})}\BibitemShut {NoStop}%
\bibitem [{\citenamefont {Anderson}\ \emph {et~al.}(2021)\citenamefont
  {Anderson}, \citenamefont {Sapiro},\ and\ \citenamefont
  {Raithel}}]{anderson2}%
  \BibitemOpen
  \bibfield  {author} {\bibinfo {author} {\bibfnamefont {D.~A.}\ \bibnamefont
  {Anderson}}, \bibinfo {author} {\bibfnamefont {R.~E.}\ \bibnamefont
  {Sapiro}},\ and\ \bibinfo {author} {\bibfnamefont {G.}~\bibnamefont
  {Raithel}},\ }\bibfield  {title} {\bibinfo {title} {An atomic receiver for am
  and fm radio communication},\ }\href
  {https://doi.org/10.1109/TAP.2020.2987112} {\bibfield  {journal} {\bibinfo
  {journal} {IEEE Transactions on Antennas and Propagation}\ }\textbf {\bibinfo
  {volume} {69}},\ \bibinfo {pages} {2455} (\bibinfo {year}
  {2021})}\BibitemShut {NoStop}%
\bibitem [{\citenamefont {Otto}\ \emph {et~al.}(2021)\citenamefont {Otto},
  \citenamefont {Hunter}, \citenamefont {Kjærgaard},\ and\ \citenamefont
  {Deb}}]{Otto21}%
  \BibitemOpen
  \bibfield  {author} {\bibinfo {author} {\bibfnamefont {J.~S.}\ \bibnamefont
  {Otto}}, \bibinfo {author} {\bibfnamefont {M.~K.}\ \bibnamefont {Hunter}},
  \bibinfo {author} {\bibfnamefont {N.}~\bibnamefont {Kjærgaard}},\ and\
  \bibinfo {author} {\bibfnamefont {A.~B.}\ \bibnamefont {Deb}},\ }\bibfield
  {title} {\bibinfo {title} {{Data capacity scaling of a distributed Rydberg
  atomic receiver array}},\ }\href {https://doi.org/10.1063/5.0048415}
  {\bibfield  {journal} {\bibinfo  {journal} {Journal of Applied Physics}\
  }\textbf {\bibinfo {volume} {129}},\ \bibinfo {pages} {154503} (\bibinfo
  {year} {2021})},\ \Eprint
  {https://arxiv.org/abs/https://pubs.aip.org/aip/jap/article-pdf/doi/10.1063/5.0048415/19763810/154503\_1\_online.pdf}
  {https://pubs.aip.org/aip/jap/article-pdf/doi/10.1063/5.0048415/19763810/154503\_1\_online.pdf}
  \BibitemShut {NoStop}%
\bibitem [{\citenamefont {Simons}\ \emph {et~al.}(2019)\citenamefont {Simons},
  \citenamefont {Haddab}, \citenamefont {Gordon},\ and\ \citenamefont
  {Holloway}}]{simons2019rydberg}%
  \BibitemOpen
  \bibfield  {author} {\bibinfo {author} {\bibfnamefont {M.~T.}\ \bibnamefont
  {Simons}}, \bibinfo {author} {\bibfnamefont {A.~H.}\ \bibnamefont {Haddab}},
  \bibinfo {author} {\bibfnamefont {J.~A.}\ \bibnamefont {Gordon}},\ and\
  \bibinfo {author} {\bibfnamefont {C.~L.}\ \bibnamefont {Holloway}},\
  }\bibfield  {title} {\bibinfo {title} {A rydberg atom-based mixer: Measuring
  the phase of a radio frequency wave},\ }\href@noop {} {\bibfield  {journal}
  {\bibinfo  {journal} {Applied Physics Letters}\ }\textbf {\bibinfo {volume}
  {114}} (\bibinfo {year} {2019})}\BibitemShut {NoStop}%
\bibitem [{\citenamefont {Jing}\ \emph {et~al.}(2020)\citenamefont {Jing},
  \citenamefont {Hu}, \citenamefont {Ma}, \citenamefont {Zhang}, \citenamefont
  {Zhang}, \citenamefont {Xiao},\ and\ \citenamefont {Jia}}]{jing2020atomic}%
  \BibitemOpen
  \bibfield  {author} {\bibinfo {author} {\bibfnamefont {M.}~\bibnamefont
  {Jing}}, \bibinfo {author} {\bibfnamefont {Y.}~\bibnamefont {Hu}}, \bibinfo
  {author} {\bibfnamefont {J.}~\bibnamefont {Ma}}, \bibinfo {author}
  {\bibfnamefont {H.}~\bibnamefont {Zhang}}, \bibinfo {author} {\bibfnamefont
  {L.}~\bibnamefont {Zhang}}, \bibinfo {author} {\bibfnamefont
  {L.}~\bibnamefont {Xiao}},\ and\ \bibinfo {author} {\bibfnamefont
  {S.}~\bibnamefont {Jia}},\ }\bibfield  {title} {\bibinfo {title} {Atomic
  superheterodyne receiver based on microwave-dressed rydberg spectroscopy},\
  }\href@noop {} {\bibfield  {journal} {\bibinfo  {journal} {Nature Physics}\
  }\textbf {\bibinfo {volume} {16}},\ \bibinfo {pages} {911} (\bibinfo {year}
  {2020})}\BibitemShut {NoStop}%
\bibitem [{\citenamefont {Firdoshi}\ \emph {et~al.}(2022)\citenamefont
  {Firdoshi}, \citenamefont {Garain}, \citenamefont {Mondal},\ and\
  \citenamefont {Mohapatra}}]{firdoshi2022six}%
  \BibitemOpen
  \bibfield  {author} {\bibinfo {author} {\bibfnamefont {T.}~\bibnamefont
  {Firdoshi}}, \bibinfo {author} {\bibfnamefont {S.}~\bibnamefont {Garain}},
  \bibinfo {author} {\bibfnamefont {S.}~\bibnamefont {Mondal}},\ and\ \bibinfo
  {author} {\bibfnamefont {A.~K.}\ \bibnamefont {Mohapatra}},\ }\bibfield
  {title} {\bibinfo {title} {Six-wave mixing of optical and microwave fields
  using rydberg excitations in thermal atomic vapor},\ }\href@noop {}
  {\bibfield  {journal} {\bibinfo  {journal} {arXiv preprint arXiv:2207.01835}\
  } (\bibinfo {year} {2022})}\BibitemShut {NoStop}%
\bibitem [{\citenamefont {Yang}\ \emph {et~al.}(2024)\citenamefont {Yang},
  \citenamefont {Yan}, \citenamefont {Li}, \citenamefont {Xiao}, \citenamefont
  {Li}, \citenamefont {Chen}, \citenamefont {Deng},\ and\ \citenamefont
  {Cheng}}]{yang2023high}%
  \BibitemOpen
  \bibfield  {author} {\bibinfo {author} {\bibfnamefont {B.}~\bibnamefont
  {Yang}}, \bibinfo {author} {\bibfnamefont {Y.}~\bibnamefont {Yan}}, \bibinfo
  {author} {\bibfnamefont {X.}~\bibnamefont {Li}}, \bibinfo {author}
  {\bibfnamefont {L.}~\bibnamefont {Xiao}}, \bibinfo {author} {\bibfnamefont
  {X.}~\bibnamefont {Li}}, \bibinfo {author} {\bibfnamefont {L.}~\bibnamefont
  {Chen}}, \bibinfo {author} {\bibfnamefont {J.}~\bibnamefont {Deng}},\ and\
  \bibinfo {author} {\bibfnamefont {H.}~\bibnamefont {Cheng}},\ }\bibfield
  {title} {\bibinfo {title} {Highly sensitive microwave electrometry with
  enhanced instantaneous bandwidth},\ }\href
  {https://doi.org/10.1103/PhysRevApplied.21.L031003} {\bibfield  {journal}
  {\bibinfo  {journal} {Phys. Rev. Appl.}\ }\textbf {\bibinfo {volume} {21}},\
  \bibinfo {pages} {L031003} (\bibinfo {year} {2024})}\BibitemShut {NoStop}%
\bibitem [{\citenamefont {Safronova}\ and\ \citenamefont
  {Safronova}(2011)}]{SAF11}%
  \BibitemOpen
  \bibfield  {author} {\bibinfo {author} {\bibfnamefont {M.~S.}\ \bibnamefont
  {Safronova}}\ and\ \bibinfo {author} {\bibfnamefont {U.~I.}\ \bibnamefont
  {Safronova}},\ }\bibfield  {title} {\bibinfo {title} {Critically evaluated
  theoretical energies, lifetimes, hyperfine constants, and multipole
  polarizabilities in $^{87}\mathrm{Rb}$},\ }\href
  {https://doi.org/10.1103/PhysRevA.83.052508} {\bibfield  {journal} {\bibinfo
  {journal} {Phys. Rev. A}\ }\textbf {\bibinfo {volume} {83}},\ \bibinfo
  {pages} {052508} (\bibinfo {year} {2011})}\BibitemShut {NoStop}%
\bibitem [{\citenamefont {Beterov}\ \emph {et~al.}(2009)\citenamefont
  {Beterov}, \citenamefont {Ryabtsev}, \citenamefont {Tretyakov},\ and\
  \citenamefont {Entin}}]{BET09}%
  \BibitemOpen
  \bibfield  {author} {\bibinfo {author} {\bibfnamefont {I.~I.}\ \bibnamefont
  {Beterov}}, \bibinfo {author} {\bibfnamefont {I.~I.}\ \bibnamefont
  {Ryabtsev}}, \bibinfo {author} {\bibfnamefont {D.~B.}\ \bibnamefont
  {Tretyakov}},\ and\ \bibinfo {author} {\bibfnamefont {V.~M.}\ \bibnamefont
  {Entin}},\ }\bibfield  {title} {\bibinfo {title} {Quasiclassical calculations
  of blackbody-radiation-induced depopulation rates and effective lifetimes of
  rydberg $ns$, $np$, and $nd$ alkali-metal atoms with $n\ensuremath{\le}80$},\
  }\href {https://doi.org/10.1103/PhysRevA.79.052504} {\bibfield  {journal}
  {\bibinfo  {journal} {Phys. Rev. A}\ }\textbf {\bibinfo {volume} {79}},\
  \bibinfo {pages} {052504} (\bibinfo {year} {2009})}\BibitemShut {NoStop}%
\bibitem [{\citenamefont {Meyer}\ \emph
  {et~al.}(2018{\natexlab{b}})\citenamefont {Meyer}, \citenamefont {Cox},
  \citenamefont {Fatemi},\ and\ \citenamefont {Kunz}}]{meyer2018digital}%
  \BibitemOpen
  \bibfield  {author} {\bibinfo {author} {\bibfnamefont {D.~H.}\ \bibnamefont
  {Meyer}}, \bibinfo {author} {\bibfnamefont {K.~C.}\ \bibnamefont {Cox}},
  \bibinfo {author} {\bibfnamefont {F.~K.}\ \bibnamefont {Fatemi}},\ and\
  \bibinfo {author} {\bibfnamefont {P.~D.}\ \bibnamefont {Kunz}},\ }\bibfield
  {title} {\bibinfo {title} {Digital communication with rydberg atoms and
  amplitude-modulated microwave fields},\ }\href@noop {} {\bibfield  {journal}
  {\bibinfo  {journal} {Applied Physics Letters}\ }\textbf {\bibinfo {volume}
  {112}} (\bibinfo {year} {2018}{\natexlab{b}})}\BibitemShut {NoStop}%
\bibitem [{\citenamefont {Hu}\ \emph {et~al.}(2023)\citenamefont {Hu},
  \citenamefont {Jiao}, \citenamefont {He}, \citenamefont {Zhang},
  \citenamefont {Zhang}, \citenamefont {Zhao},\ and\ \citenamefont
  {Jia}}]{Hu2023}%
  \BibitemOpen
  \bibfield  {author} {\bibinfo {author} {\bibfnamefont {J.}~\bibnamefont
  {Hu}}, \bibinfo {author} {\bibfnamefont {Y.}~\bibnamefont {Jiao}}, \bibinfo
  {author} {\bibfnamefont {Y.}~\bibnamefont {He}}, \bibinfo {author}
  {\bibfnamefont {H.}~\bibnamefont {Zhang}}, \bibinfo {author} {\bibfnamefont
  {L.}~\bibnamefont {Zhang}}, \bibinfo {author} {\bibfnamefont
  {J.}~\bibnamefont {Zhao}},\ and\ \bibinfo {author} {\bibfnamefont
  {S.}~\bibnamefont {Jia}},\ }\bibfield  {title} {\bibinfo {title} {Improvement
  of response bandwidth and sensitivity of rydberg receiver using multi-channel
  excitations},\ }\href {https://doi.org/10.1140/epjqt/s40507-023-00209-7}
  {\bibfield  {journal} {\bibinfo  {journal} {EPJ Quantum Technology}\ }\textbf
  {\bibinfo {volume} {10}},\ \bibinfo {pages} {51} (\bibinfo {year}
  {2023})}\BibitemShut {NoStop}%
\bibitem [{\citenamefont {Artusio-Glimpse}\ \emph {et~al.}(2024)\citenamefont
  {Artusio-Glimpse}, \citenamefont {Long}, \citenamefont {Bresler},
  \citenamefont {Prajapati}, \citenamefont {Shylla}, \citenamefont {Rotunno},
  \citenamefont {Simons}, \citenamefont {Berweger}, \citenamefont
  {Schlossberger}, \citenamefont {Lebrun},\ and\ \citenamefont
  {Holloway}}]{AlyFCIB24}%
  \BibitemOpen
  \bibfield  {author} {\bibinfo {author} {\bibfnamefont {A.~B.}\ \bibnamefont
  {Artusio-Glimpse}}, \bibinfo {author} {\bibfnamefont {D.~A.}\ \bibnamefont
  {Long}}, \bibinfo {author} {\bibfnamefont {S.~M.}\ \bibnamefont {Bresler}},
  \bibinfo {author} {\bibfnamefont {N.}~\bibnamefont {Prajapati}}, \bibinfo
  {author} {\bibfnamefont {D.}~\bibnamefont {Shylla}}, \bibinfo {author}
  {\bibfnamefont {A.~P.}\ \bibnamefont {Rotunno}}, \bibinfo {author}
  {\bibfnamefont {M.~T.}\ \bibnamefont {Simons}}, \bibinfo {author}
  {\bibfnamefont {S.}~\bibnamefont {Berweger}}, \bibinfo {author}
  {\bibfnamefont {N.}~\bibnamefont {Schlossberger}}, \bibinfo {author}
  {\bibfnamefont {T.~W.}\ \bibnamefont {Lebrun}},\ and\ \bibinfo {author}
  {\bibfnamefont {C.~L.}\ \bibnamefont {Holloway}},\ }\bibfield  {title}
  {\bibinfo {title} {Increased instantaneous bandwidth of rydberg atom
  electrometry with an optical frequency comb probe},\ }\href
  {arXiv:2402.17942} {\  (\bibinfo {year} {2024})}\BibitemShut {NoStop}%
\bibitem [{\citenamefont {Bohaichuk}\ \emph {et~al.}(2022)\citenamefont
  {Bohaichuk}, \citenamefont {Booth}, \citenamefont {Nickerson}, \citenamefont
  {Tai},\ and\ \citenamefont {Shaffer}}]{bohaichuk2022origins}%
  \BibitemOpen
  \bibfield  {author} {\bibinfo {author} {\bibfnamefont {S.~M.}\ \bibnamefont
  {Bohaichuk}}, \bibinfo {author} {\bibfnamefont {D.}~\bibnamefont {Booth}},
  \bibinfo {author} {\bibfnamefont {K.}~\bibnamefont {Nickerson}}, \bibinfo
  {author} {\bibfnamefont {H.}~\bibnamefont {Tai}},\ and\ \bibinfo {author}
  {\bibfnamefont {J.~P.}\ \bibnamefont {Shaffer}},\ }\bibfield  {title}
  {\bibinfo {title} {Origins of rydberg-atom electrometer transient response
  and its impact on radio-frequency pulse sensing},\ }\href@noop {} {\bibfield
  {journal} {\bibinfo  {journal} {Physical Review Applied}\ }\textbf {\bibinfo
  {volume} {18}},\ \bibinfo {pages} {034030} (\bibinfo {year}
  {2022})}\BibitemShut {NoStop}%
\end{thebibliography}%

\end{document}